\documentclass[aps,prl,twocolumn,superscriptaddress,showpacs,epsfig]{revtex4}
\usepackage{graphicx}
\usepackage{graphics}
\usepackage{epsfig}
\usepackage{subfigure}
\usepackage{longtable}
\newcommand{\met} {\hbox{$E$\kern-0.55em\lower-0.15ex\hbox{/}}_T}
\newcommand{\vetmet} {\hbox{$\vec{E}$\kern-0.55em\lower-0.15ex\hbox{/}}_T}
\begin{document}

\title{Measurement of the top-quark mass using missing $E_T$+jets
events with secondary vertex $b$-tagging at CDF II}

\affiliation{Institute of Physics, Academia Sinica, Taipei, Taiwan 11529, Republic of China} 
\affiliation{Argonne National Laboratory, Argonne, Illinois 60439} 
\affiliation{Institut de Fisica d'Altes Energies, Universitat Autonoma de Barcelona, E-08193, Bellaterra (Barcelona), Spain} 
\affiliation{Baylor University, Waco, Texas  76798} 
\affiliation{Istituto Nazionale di Fisica Nucleare, University of Bologna, I-40127 Bologna, Italy} 
\affiliation{Brandeis University, Waltham, Massachusetts 02254} 
\affiliation{University of California, Davis, Davis, California  95616} 
\affiliation{University of California, Los Angeles, Los Angeles, California  90024} 
\affiliation{University of California, San Diego, La Jolla, California  92093} 
\affiliation{University of California, Santa Barbara, Santa Barbara, California 93106} 
\affiliation{Instituto de Fisica de Cantabria, CSIC-University of Cantabria, 39005 Santander, Spain} 
\affiliation{Carnegie Mellon University, Pittsburgh, PA  15213} 
\affiliation{Enrico Fermi Institute, University of Chicago, Chicago, Illinois 60637} 
\affiliation{Comenius University, 842 48 Bratislava, Slovakia; Institute of Experimental Physics, 040 01 Kosice, Slovakia} 
\affiliation{Joint Institute for Nuclear Research, RU-141980 Dubna, Russia} 
\affiliation{Duke University, Durham, North Carolina  27708} 
\affiliation{Fermi National Accelerator Laboratory, Batavia, Illinois 60510} 
\affiliation{University of Florida, Gainesville, Florida  32611} 
\affiliation{Laboratori Nazionali di Frascati, Istituto Nazionale di Fisica Nucleare, I-00044 Frascati, Italy} 
\affiliation{University of Geneva, CH-1211 Geneva 4, Switzerland} 
\affiliation{Glasgow University, Glasgow G12 8QQ, United Kingdom} 
\affiliation{Harvard University, Cambridge, Massachusetts 02138} 
\affiliation{Division of High Energy Physics, Department of Physics, University of Helsinki and Helsinki Institute of Physics, FIN-00014, Helsinki, Finland} 
\affiliation{University of Illinois, Urbana, Illinois 61801} 
\affiliation{The Johns Hopkins University, Baltimore, Maryland 21218} 
\affiliation{Institut f\"{u}r Experimentelle Kernphysik, Universit\"{a}t Karlsruhe, 76128 Karlsruhe, Germany} 
\affiliation{High Energy Accelerator Research Organization (KEK), Tsukuba, Ibaraki 305, Japan} 
\affiliation{Center for High Energy Physics: Kyungpook National University, Taegu 702-701, Korea; Seoul National University, Seoul 151-742, Korea; SungKyunKwan University, Suwon 440-746, Korea} 
\affiliation{Ernest Orlando Lawrence Berkeley National Laboratory, Berkeley, California 94720} 
\affiliation{University of Liverpool, Liverpool L69 7ZE, United Kingdom} 
\affiliation{University College London, London WC1E 6BT, United Kingdom} 
\affiliation{Centro de Investigaciones Energeticas Medioambientales y Tecnologicas, E-28040 Madrid, Spain} 
\affiliation{Massachusetts Institute of Technology, Cambridge, Massachusetts  02139} 
\affiliation{Institute of Particle Physics: McGill University, Montr\'{e}al, Canada H3A~2T8; and University of Toronto, Toronto, Canada M5S~1A7} 
\affiliation{University of Michigan, Ann Arbor, Michigan 48109} 
\affiliation{Michigan State University, East Lansing, Michigan  48824} 
\affiliation{University of New Mexico, Albuquerque, New Mexico 87131} 
\affiliation{Northwestern University, Evanston, Illinois  60208} 
\affiliation{The Ohio State University, Columbus, Ohio  43210} 
\affiliation{Okayama University, Okayama 700-8530, Japan} 
\affiliation{Osaka City University, Osaka 588, Japan} 
\affiliation{University of Oxford, Oxford OX1 3RH, United Kingdom} 
\affiliation{University of Padova, Istituto Nazionale di Fisica Nucleare, Sezione di Padova-Trento, I-35131 Padova, Italy} 
\affiliation{LPNHE, Universite Pierre et Marie Curie/IN2P3-CNRS, UMR7585, Paris, F-75252 France} 
\affiliation{University of Pennsylvania, Philadelphia, Pennsylvania 19104} 
\affiliation{Istituto Nazionale di Fisica Nucleare Pisa, Universities of Pisa, Siena and Scuola Normale Superiore, I-56127 Pisa, Italy} 
\affiliation{University of Pittsburgh, Pittsburgh, Pennsylvania 15260} 
\affiliation{Purdue University, West Lafayette, Indiana 47907} 
\affiliation{University of Rochester, Rochester, New York 14627} 
\affiliation{The Rockefeller University, New York, New York 10021} 
\affiliation{Istituto Nazionale di Fisica Nucleare, Sezione di Roma 1, University of Rome ``La Sapienza," I-00185 Roma, Italy} 
\affiliation{Rutgers University, Piscataway, New Jersey 08855} 
\affiliation{Texas A\&M University, College Station, Texas 77843} 
\affiliation{Istituto Nazionale di Fisica Nucleare, University of Trieste/\ Udine, Italy} 
\affiliation{University of Tsukuba, Tsukuba, Ibaraki 305, Japan} 
\affiliation{Tufts University, Medford, Massachusetts 02155} 
\affiliation{Waseda University, Tokyo 169, Japan} 
\affiliation{Wayne State University, Detroit, Michigan  48201} 
\affiliation{University of Wisconsin, Madison, Wisconsin 53706} 
\affiliation{Yale University, New Haven, Connecticut 06520} 
\author{T.~Aaltonen}
\affiliation{Division of High Energy Physics, Department of Physics, University of Helsinki and Helsinki Institute of Physics, FIN-00014, Helsinki, Finland}
\author{A.~Abulencia}
\affiliation{University of Illinois, Urbana, Illinois 61801}
\author{J.~Adelman}
\affiliation{Enrico Fermi Institute, University of Chicago, Chicago, Illinois 60637}
\author{T.~Affolder}
\affiliation{University of California, Santa Barbara, Santa Barbara, California 93106}
\author{T.~Akimoto}
\affiliation{University of Tsukuba, Tsukuba, Ibaraki 305, Japan}
\author{M.G.~Albrow}
\affiliation{Fermi National Accelerator Laboratory, Batavia, Illinois 60510}
\author{S.~Amerio}
\affiliation{University of Padova, Istituto Nazionale di Fisica Nucleare, Sezione di Padova-Trento, I-35131 Padova, Italy}
\author{D.~Amidei}
\affiliation{University of Michigan, Ann Arbor, Michigan 48109}
\author{A.~Anastassov}
\affiliation{Rutgers University, Piscataway, New Jersey 08855}
\author{K.~Anikeev}
\affiliation{Fermi National Accelerator Laboratory, Batavia, Illinois 60510}
% ...
\collaboration{CDF Collaboration\footnote{With visitors from $^a$University of Athens, 
$^b$University of Bristol, 
$^c$University Libre de Bruxelles, 
$^d$Cornell University, 
$^e$University of Cyprus, 
$^f$University of Dublin, 
$^g$University of Edinburgh, 
$^h$University of Heidelberg, 
$^i$Universidad Iberoamericana, 
$^j$University of Manchester, 
$^k$Nagasaki Institute of Applied Science, 
$^l$University de Oviedo, 
$^m$University of London, Queen Mary College, 
$^n$University of California Santa Cruz, 
$^o$Texas Tech University, 
$^p$University of California Irvine, 
$^q$IFIC(CSIC-Universitat de Valencia). 
}}
\noaffiliation

\begin{abstract}
\noindent
We present a measurement of the top-quark mass in $p\bar p$ collisions at 
\hbox{$\sqrt{s}=1.96$} TeV which uses events with an inclusive signature of 
missing transverse energy and jets. 
The event selection is sensitive to $t\bar t\to W^+bW^-\bar b\to l\nu b
q q^\prime \bar b$ independent of the lepton flavor, and results in 
a large acceptance for $W\to\tau\nu$ decays.
All-hadronic $t\bar t$ decays and events with identified electrons or muons are 
vetoed to provide a statistically independent sample with respect to 
all previous measurements. 
The top-quark mass is inferred from the distribution of the scalar sum
of all jet transverse energies and the missing transverse energy.
Using $311$ pb$^{-1}$ of integrated luminosity recorded by the Collider Detector 
at Fermilab, we measure a top-quark mass  
$m_t = 172.3 ^{+10.8}_{-9.6}$(stat.)$\pm 10.8$(syst.) GeV/$c^2$.
While the uncertainty on $m_t$ is larger than that of other measurements, 
the result is statistically uncorrelated with those of other methods, and 
thus can help to reduce the overall $m_t$ uncertainty when
combined with other existing measurements.  
\end{abstract}

\maketitle


%
%
\noindent

The top-quark mass, $m_t$, is an important free parameter in the Standard Model (SM) of particle
physics. Being roughly 40 times larger than the mass of its weak isospin partner, the $b$ quark,
$m_t$ gives large contributions to electroweak radiative corrections which,  when 
connected to precision electroweak measurements, can 
be used to derive constraints on the masses of the yet-unobserved Higgs boson~\cite{higgslim}, and of particles belonging to some SM extensions~\cite{extlim}.
At the Tevatron $p\bar p$ collider top quarks are produced mainly in pairs
through quark-antiquark annihilation and gluon-gluon fusion processes.
Because the Cabibbo-Kobayashi-Maskawa matrix element $V_{tb}$~\cite{PDG} is close
to unity, the SM top quark decays  to a $W$ boson 
and a $b$ quark almost $100\%$ of the time. The final state of a top-quark pair 
thus includes two $W$ bosons and two $b$-quark jets. 
When only one $W$ decays leptonically, the $t\bar t$ event typically contains a charged 
lepton, missing transverse energy ($\met$)  from the undetected neutrino~\cite{CDFcoord}, 
and four high-transverse-energy jets, two of which originate from $b$ quarks.

Recently the CDF collaboration has reported precision $m_t$ measurements using
$t\bar t$ events containing identified high-$p_T$ leptons ($e$, $\mu$)~\cite{lj_mass}
and all-hadronic decays~\cite{allhad_mass}. 
In this paper we describe a top-quark mass measurement which uses events 
collected by a multijet trigger, and selected by requiring an inclusive 
high-$p_T$ neutrino signature, consisting of large $\met$. 
Events containing identified high-$p_T$ electrons or muons ($E_T^{e} \ge 20$
GeV, $P_T^\mu \ge 20$ GeV/$c$),
as defined in~\cite{cdfxsec2secvtx}, are removed in order to increase the
relative contribution of $W\to \tau \nu$ decays and provide a 
statistically independent sample with respect to other lepton-based 
measurements~\cite{lj_mass}. All-hadronic $t\bar t$ decays are discarded 
by the $\met$ requirement so that orthogonality with respect to the all-hadronic
mass sample is ensured~\cite{allhad_mass}.
Unlike previous analyses based on the identification of $W\to e\nu (\mu \nu)$ and
$W\to qq^\prime$ 
decays, our event selection does not permit a full kinematical reconstruction 
of the $t\bar t$ final state. For this reason, 
the top-quark mass is derived  from the $H_T$ distribution, where $H_T$ is
defined as the scalar sum of all jet transverse energies and the $\met$. 
%

%
%
%

Results reported in this paper  are obtained using $311$ pb$^{-1}$ 
of integrated luminosity from $p\bar p$ collisions at $\sqrt{s} =1.96$ TeV, 
recorded by the Collider Detector at Fermilab (CDF\,II).
The CDF\,II detector is described in detail elsewhere~\cite{CDFII}. 
It consists of a magnetic spectrometer surrounded by 
a calorimeter and muon system. The momenta of charged particles 
are measured up to a pseudorapidity of $|\eta|=1.0$ in a cylindrical
drift chamber, which is inside a 1.4 T superconducting solenoidal magnet. 
Silicon microstrip vertex detectors, located immediately outside the beampipe,
provide precise track reconstruction useful for vertexing and 
extend the coverage of the tracking system up to $|\eta|=2.0$.
Electromagnetic and hadronic sampling calorimeters, arranged in a 
projective-tower geometry, surround the tracking systems and measure the 
energy and direction of electrons, photons, and jets in the 
range $|\eta|<3.6$. In addition, the good hermeticity provided by 
the calorimeter allows the detection of high-$p_T$ neutrinos by 
the measurement of the $\met$.
Muon systems outside the calorimeters allow the reconstruction of 
track segments for penetrating particles.
The beam luminosity is determined using gas Cherenkov counters surrounding the 
beam pipe,
which measure the average number of inelastic $p\bar p$ collisions per bunch crossing. 
%
%
%

The data sample used in this analysis is collected by a multijet trigger 
which requires four or more  $E_T\ge 15$ GeV clusters  of contiguous 
calorimeter towers, and a scalar sum of transverse 
energy clustered in the calorimeter of $\sum E_T \ge 125$ GeV. 
The initial data sample consists of $4.2$ million events and is further  
reduced offline by the application of kinematical and topological requirements 
aimed at optimizing the $t\bar t$ signal significance~\cite{PRLmetjet}.
Briefly, we require at least four jets having $E_T\ge 15$ GeV and $|\eta|\le 2.0$; $\met$ significance,
$\met^{\rm sig}$, greater than $4.0$ GeV$^{1/2}$, where $\met^{\rm sig}$ is defined as 
$ \met / \sqrt{\sum E_T}$; and a minimum separation in azimuthal angle 
between the $\met$ and the closest jet, $\min\Delta\phi \ge 0.4$ rad. 
In our selection, 
jets are identified as groups of calorimeter tower energy deposits within a cone of radius 
$\Delta R = \sqrt{\Delta \phi^2 + \Delta\eta^2} \le 0.4$, and their energies are 
corrected for calorimeter  non-linearity, losses in the gaps between 
towers, multiple interactions, and particle response calibrations~\cite{JetCorr}.
This selection reduces the data sample to $597$ events, with a signal 
to background ratio $S/B\sim 1/5$. 
In order to further increase the expected $S/B$ ratio and reject background events 
with only light quark or gluon jets,  $b$-quark jets (``$b$ tags'') are 
identified by the reconstruction of secondary decay vertices using the 
{\scshape{secvtx}} algorithm, as in~\cite{cdfxsec2secvtx}.
After these selections and the requirement of at least one $b$-tagged jet,
we observe 106 events with $S/B\sim 1$; about $44\%$ of the signal acceptance
is accounted for by $t\bar t\to W^+bW^-\bar b\to \tau\nu b
q q^\prime \bar b$ decays, while
the remaining $t\bar t$ content is dominated by
$e(\mu)$+jets events, in which the lepton fails the standard 
high-$p_T$ identification cuts.
%
%
%
%

Background events with $b$ tags arise from QCD heavy flavor production, 
electroweak production of $W$ bosons associated with heavy flavor jets, 
and from false identification by the {\scshape{secvtx}} algorithm.
%
%
%
\begin{figure}[Ht!]
\begin{center}
\epsfig{figure=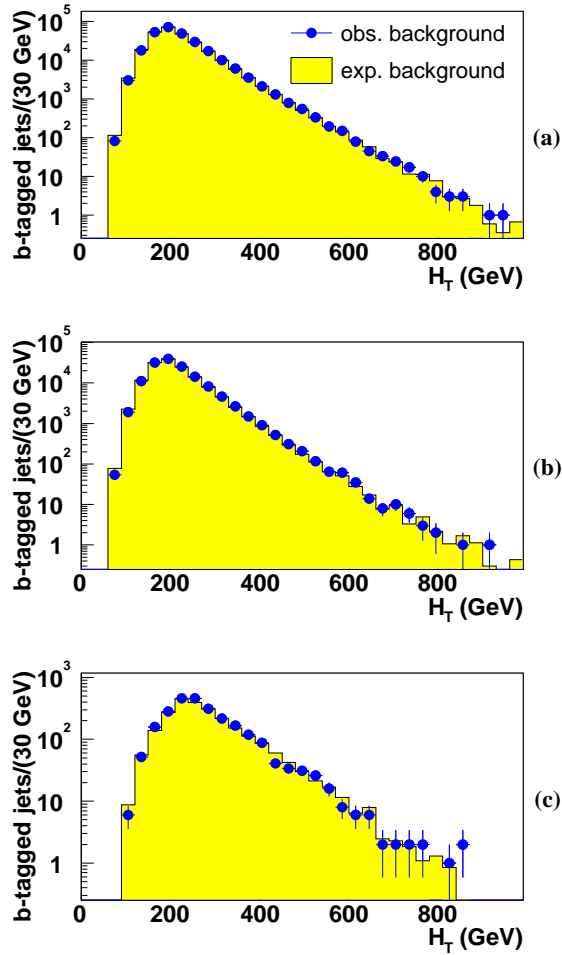,
                     width=.9\linewidth,clip=}
\caption{Observed and expected $H_T$ background distributions in data control
         samples depleted of signal contamination, see text for details.}
\label{f:mtxcheck_ht}
\end{center}
\end{figure}
%
%
%
The overall number of background $b$ tags in the final data sample, 
and their corresponding kinematical distributions, are estimated using
a per-jet parameterization of the $b$-tagging probability derived from 
the multijet sample. For the parameterization, 
we use events with exactly three jets, having $E_T\ge 15$ GeV and $|\eta|\le 1.0$,
where the $t\bar t$ content is negligible.
The parameterization exploits the $b$-tag rate dependencies on the jet $E_T$, 
the charged track multiplicity inside the jet cone, and the projection of the 
$\met$ along the jet direction in the transverse plane, which is defined by 
$\met^{\rm prj} = \met \cos\Delta\phi(\met,{\rm jet})$.
The extrapolation of the 3-jet $b$-tagging probability to higher jet multiplicity
events, and the capability of the parameterization to track sample 
composition changes introduced by the kinematical selection, are checked
using $\ge\!4$-jet data samples depleted of signal content, as described 
elsewhere~\cite{PRLmetjet}:
(a) data before the optimized kinematical selection on $\met^{\rm sig}$ and 
$\min\Delta\phi(\met,{\rm jets})$; 
(b) $\met^{\rm sig} \le 3.0$ GeV$^{1/2}$, $\min\Delta\phi(\met,{\rm jets})\ge 0.3$ rad, and 
(c) $\met^{\rm sig} \ge 3.0$ GeV$^{1/2}$, $\min\Delta\phi(\met,{\rm jets})\le 0.3$ rad.
As a result, the $b$-tag rate parameterization is found to predict the 
number of background $b$ tags, and the shape of their corresponding kinematical 
distributions, to within $10\%$ in the $4\le N_{\rm jet} \le 6$
region, where $96.4\%$ of the $t\bar t$ signal is expected after the optimized 
kinematical selection. 
Figure~\ref{f:mtxcheck_ht} shows the comparison between
expected and observed background $H_T$ distributions in the
data control samples (a), (b), and (c). 
The expected $H_T$ distributions are derived from the $b$-tag rate 
parameterization applied to each jet belonging to a given  data sample, before 
$b$-jet identification requirements. The observed $H_T$ distributions 
receive one entry per $b$-tagged jet for a proper normalization with
the expectation. The normalization  and shape of the observed and expected 
distributions are in good agreement for all control samples. 
%
%
%

The final data sample, after the optimized kinematical selection and the additional requirement of at 
least one  $b$-tagged jet, contains a total of $127$ $b$-tagged jets. 
The number of $b$-tagged jets yielded by background processes in that sample is expected to be 
$n_b^{\rm exp} = 57.4\pm 8.1$. The excess in the number of $b$ tags is ascribed to top-quark pair production. 
%
%
We derive a measurement of the  top-quark mass from the observed $H_T$ distribution.
%
The $H_T$ distribution from the selected data
is fit to the sum of signal and background $H_T$ contribution
parameterizations using an unbinned likelihood technique.
Probability density functions 
are determined for signal, as a function of $m_t$,  and for
background events by fitting a functional form from the corresponding 
$H_T$ distributions (templates).
For consistency with our per-jet background prediction method, the $H_T$ distributions
from data and simulated signal events receive one entry per $b$-tagged jet.
%
%
%
\begin{figure}[Ht!]
\begin{center}
\epsfig{figure=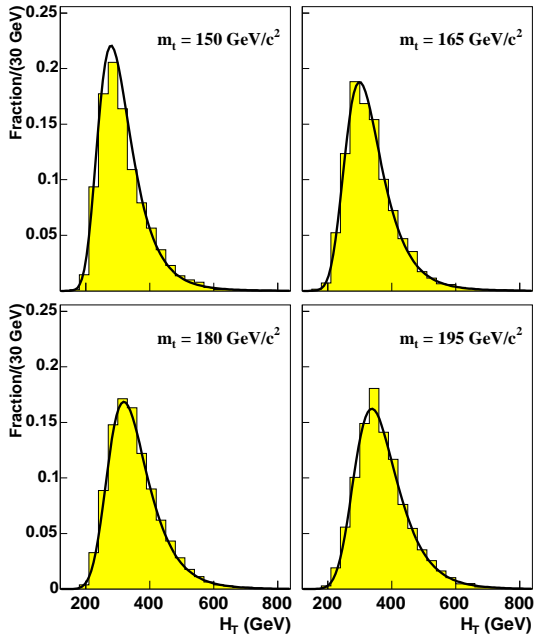,
                     width=.9\linewidth,clip=}
\caption{Four $H_T$ signal templates with $m_t$ ranging from 150 to 195 GeV/$c^2$.
         Overlaid are the fitted parameterizations at each generated top-quark mass. }
\label{f:mctemp}
\end{center}
\end{figure}
%
%
%
%
%

We calibrate our method using events with inclusive $t\bar t$ decays
generated with different input values of 
$m_t$ ranging from 150 to 200 GeV/$c^2$, in steps of 2.5 GeV/$c^2$. These events 
are simulated using the {\scshape{herwig}}~\cite{herwig} generator in conjunction with the
{\scshape{cteq5l}}~\cite{CTEQ} parton distribution functions (PDFs),
{\scshape{qq}}~\cite{QQ} for the modeling of $b$ and $c$ hadron decays,
and a full simulation of the CDF\,II detector~\cite{Gerchtein:2003ba, Geant}. They
are then subjected to the same selection as the recorded events.
%
%
The $H_T$ distributions, derived at discrete values of the 
top-quark mass, are parameterized by a continuous functional 
form as a function of $m_t$ in order to smooth the distributions and 
interpolate between the templates. 
For any given $m_t$ the probability to observe a particular $H_T$ value 
is specified by a normalized Pearson type IV function~\cite{Pearson_IV},
in which the parameters  are assumed to be 
linearly dependent on $m_t$. The best parameterization is determined 
by a simultaneous binned likelihood fit to all signal templates.
In Fig.~\ref{f:mctemp}, four signal templates are shown overlaid 
with their corresponding parameterization.

\begin{figure}[Ht!]
\begin{center}
\epsfig{figure=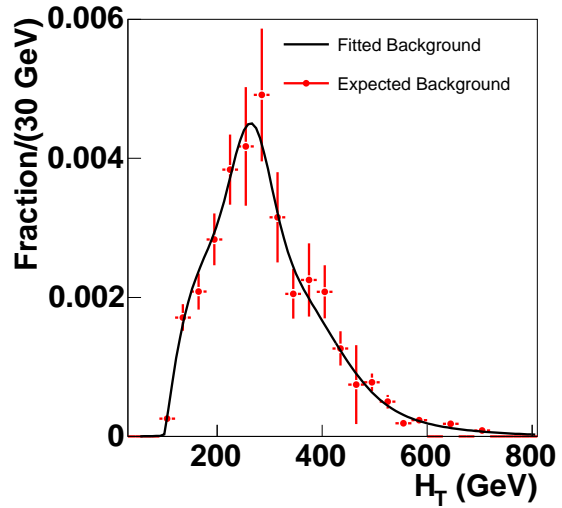,
                     width=.9\linewidth,clip=}
\caption{The background $H_T$ template, after the subtraction of the $t\bar t$  content (using $m_t$=172.5 GeV/$c^2$),
         is shown overlaid with the fitted parameterization. }
\label{f:bkgtemp}
\end{center}
\end{figure}

%
%
%
%
For background, the $H_T$ distribution is derived from the $b$-tag rate 
parameterization applied to jets belonging to the  kinematically selected data sample, 
before $b$-jet identification requirements.
It has no dependence on the top-quark mass, except from a procedure adopted to subtract 
the expected signal content
($\sim 15\%$ for $m_t = 172.5$ GeV/$c^2$). The arbitrary $m_t$ choice 
in the subtraction procedure is accounted for in the background shape systematic 
uncertainty. A single probability
density function, defined as the sum of a gamma function and 
two gaussians, is used to fit the background $H_T$ template, as 
shown in Fig.~\ref{f:bkgtemp}.
%
%
%
%
%
%
%
%
%

%
The likelihood function used to extract the top-quark mass   
includes as free parameters the number of expected signal and background 
$b$ tags ($n_s$ and $n_b$), and $m_t$. It is specified by three factors:

\begin{equation}
\label{eq:like}
\mathcal{L}(m_{t}) = \mathcal{L}_{\rm sh}(m_{t}) \times \mathcal{L}_{n_s + n_b} \times \mathcal{L}_{\rm bkg},
\end{equation}

\noindent
where

\begin{eqnarray}
\mathcal{L}_{\rm sh}(m_{t}) &=& \prod_{i=1}^N \frac{n_s \cdot P_{\rm sig}(H_T^i|m_{t}) + n_b \cdot P_{\rm bkg}(H_T^i)}{n_s + n_b}, \\ 
\label{eq:shape}
\mathcal{L}_{n_s + n_b} & = & \frac{e^{-(n_s + n_b)} \cdot (n_s + n_b)^N}{N!}, \\ 
\label{eq:num}
\mathcal{L}_{\rm bkg}       & = & e^{-\frac{1}{2}(n_b - n_b^{\rm exp})^2/\sigma_{n_b}^2}, 
\label{eq:bkg}
\end{eqnarray}

\noindent
and $N$ is the number of observed $b$ tags in the final data sample. 
In  $\mathcal{L}_{\rm sh}(m_{t})$ the product is over the number of observed $b$ tags, and
$P_{\rm sig}(H_T^i|m_{t})$ and $P_{\rm bkg}(H_T^i)$ are the probability density functions for
signal and background, respectively.
The second factor of Eq.~(\ref{eq:like}) represents a Poisson constraint on the total
number of $b$ tags observed in the data. Finally, in Eq.~(\ref{eq:bkg}) the 
background normalization is constrained to its expected value $n_b^{\rm exp}$ to within
$\sigma_{n_b}\equiv 10\% \cdot n_b^{\rm exp}$. 
The likelihood is maximized with respect to $n_s$, $n_b$ and $m_t$. 
The statistical uncertainty from the fit procedure is taken from the $m_t$ values
where the log-likelihood changes by $0.5$ units from its maximum. 
Since we are counting $b$ tags and not events, the $H_T$ distribution is not
strictly Poisson in nature. We correct for this effect below.
%
%
%
%

%
%
%

We use simulated data ensembles (pseudo-experiments) to check our fitting 
procedure for possible systematic biases. 
For each generated top-quark mass from $150$ GeV/$c^2$ to $200$ GeV/$c^2$, 
we construct pseudo-experiments, with the same statistical properties as our
observed data sample, by randomly sampling from the signal and background templates.
Then we perform likelihood fits to each pseudo-experiment and characterize the 
accuracy of the technique in determining the correct $m_t$ value. 
In each pseudo-experiment, the number of background $b$ tags is Poisson
fluctuated around its expectation, $n_b^{\rm exp}$, while the number of 
signal $b$ tags is Poisson fluctuated around the number observed in the data,
minus the central value for the background expectation. In this procedure,
$b$ tags from single and double $b$-tagged events are fluctuated separately. 
For each pseudo-experiment, the likelihood fit provides the measured $m_t$
along with the positive and negative statistical uncertainties from which
pull distributions are derived. 
The mean of the pull distribution, averaged as a function of the input $m_t$,
is consistent with zero ($-0.01 \pm 0.02)$, while the width is slightly larger 
than unity, due to the inclusion of duplicated $H_T$ values in the pseudo-experiment 
distributions in the case of double-tagged events. For the current analysis, 
we correct for this effect by scaling the statistical errors taken from 
$\Delta \ln \mathcal{L} = -1/2$. The scale factor is the pull width averaged over 
$m_t$ ranging between 150 and 200 GeV/$c^2$, giving $1.08\pm 0.02$.
%
%
%

\begin{figure}[Ht!]
\begin{center}
\epsfig{figure=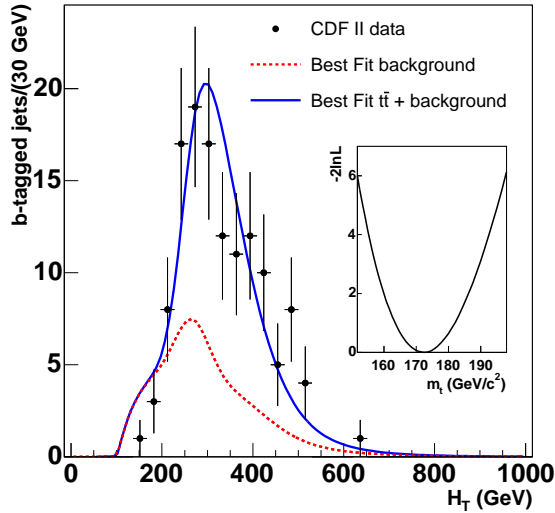,
                     width=.9\linewidth,clip=}
\caption{$H_T$ distribution from the selected data sample, overlaid with
         the expected distribution from the unbinned likelihood fit. The 
         inset shows the $-2 \ln \mathcal{L}$ from the final fit 
         as a function of $m_t$. }
\label{f:fits1}
\end{center}
\end{figure}

Applying our method to the observed $H_T$ distribution, we find
$n_s = 76.2\pm 11.4$, $n_b = 54.6\pm 5.1$, and 
$m_t = 172.3 ^{+10.8}_{-9.6}$(stat.) GeV/$c^2$. The statistical uncertainties 
on $m_t$ are consistent with expectation from pseudo-experiments performed 
with an input top-quark mass of $172.5$ GeV/$c^2$.
 The result
from the fit to the data is shown in Fig.~\ref{f:fits1}. The 
inset shows the function $-2 \ln \mathcal{L}$ from the final fit as a function of 
$m_t$.
%
%
%

%
%
%
\begin{table}
\begin{center}
\begin{ruledtabular}
\caption{Relevant sources of systematic uncertainty.}
\label{t:systsall}

\begin{tabular}{lc}
Source & $\Delta m_t$ (GeV/$c^2$) \\
\hline
Jet energy scale    &  9.6 \\
Generator           &  3.8 \\
Background shape    &  2.1 \\ 
PDFs                &  1.5 \\
ISR                 &  0.9 \\
FSR                 &  0.9 \\
Background fraction &  0.8 \\
$b$-jet energy scale  &  0.7 \\
Trigger efficiency  &  0.7 \\
Limited Monte Carlo statistics       &  0.6 \\
$b$ tagging               &  0.5 \\
\hline
Total               & 10.8 \\
\end{tabular}
\end{ruledtabular}
\end{center}
\end{table}
%
%

Systematic uncertainties arise from uncertainties in our understanding 
of the detector response and in the assumptions employed to infer the 
top-quark mass from the observed data. 
For each source of systematic 
uncertainty, the relevant input quantities are varied by 
$\pm 1 \sigma$, and new signal or background $H_T$ templates
are produced by performing the event selection and reconstruction on
the modified samples. Then these new fixed templates are used to run pseudo-experiments. 
The mean shift in the fitted top-quark mass with respect to the input value  
is taken as the systematic uncertainty associated with the given 
assumption or effect. Table~\ref{t:systsall} reports all the relevant
sources of systematics associated with our measurement. 
The dominant source of uncertainty (9.6 GeV/$c^2$) given the 
choice of $H_T$ as discriminant variable is associated to jet energy 
scale (JES) uncertainty. 
For each jet considered in the $H_T$ calculation the relative jet energy
scale  uncertainty, which is mainly driven by uncertainties on particle response 
calibrations and the out-of-cone jet energy modeling, varies from 3\% to 8\% 
depending on $\eta$ and $p_T$ of the jet. We determine the impact of the JES
uncertainty on our measurement using pseudo-experiments 
in which the nominal jet energies are varied by $\pm 1$ standard deviations. 
Additionally, the dependence on the Monte Carlo generator is estimated 
as the difference in the extracted top-quark mass in {\scshape{pythia}}~\cite{pythia} 
and {\scshape{herwig}} events, and amounts to 3.8 GeV/$c^2$.
Other sources  of uncertainty are related to the background
shape and normalization and are evaluated to be 2.1 and 0.8  GeV/$c^2$,
respectively.
We estimate the uncertainty from PDFs
using signal samples in which the events are weighted according to their 
probability to occur using different sets of PDF eigenvectors.
The systematic uncertainty is computed by considering differences between
the {\scshape{cteq5l}} and {\scshape{mrst72}}~\cite{mrst72} PDFs parameterizations, 
different $\Lambda_{QCD}$ values, 
and the sum in quadrature of half the difference between
the $\pm 1 \sigma$ shift of the 20 {\scshape{cteq6m}} uncertainties, for a total of
1.5 GeV/$c^2$.
Variation of initial ({\scshape{isr}}) and final state ({\scshape{fsr}}) gluon 
radiation settings, as in~\cite{lj_mass}, are found to contribute 0.9 GeV/$c^2$ 
of systematic uncertainty each.
Systematic uncertainties due to  the $b$-jet energy scale, 
trigger simulation effects, statistically
limited Monte Carlo samples, and  $b$-tagging efficiency
modeling, are small and give a combined error 
of 1.2 GeV/$c^2$.
The total systematic uncertainty is estimated to 
be $10.8$ GeV/$c^2$ assuming all sources to be uncorrelated. 
%
%

In conclusion, we report the first top-quark mass measurement using inclusively
selected $\met$+jets $t\bar t$  events with  a large acceptance for
$W\to\tau\nu$ decays.
The result, \hbox{$m_t = 172.3 ^{+10.8}_{-9.6}$(stat.)$\pm 10.8$(syst.) GeV/$c^2$}, is 
complementary and statistically independent with respect to precision
CDF measurements~\cite{lj_mass, allhad_mass}, and  consequently, although 
not competitive by itself,  it will help to reduce by a few percent the overall uncertainty 
on $m_t$ when combined with other existing results.

%
%
%


We thank the Fermilab staff and the technical staffs of the participating institutions for their vital contributions. This work was supported by the U.S. Department of Energy and National Science Foundation; the Italian Istituto Nazionale di Fisica Nucleare; the Ministry of Education, Culture, Sports, Science and Technology of Japan; the Natural Sciences and Engineering Research Council of Canada; the National Science Council of the Republic of China; the Swiss National Science Foundation; the A.P. Sloan Foundation; the Bundesministerium f\"ur Bildung und Forschung, Germany; the Korean Science and Engineering Foundation and the Korean Research Foundation; the Particle Physics and Astronomy Research Council and the Royal Society, UK; the Institut National de Physique Nucleaire et Physique des Particules/CNRS; the Russian Foundation for Basic Research; the Comisi\'on Interministerial de Ciencia y Tecnolog\'{\i}a, Spain; the European Community's Human Potential Programme; the Slovak R\&D Agency; and the Academy of Finland.

%
%
%

%


\end{document}